# SPEECH NEUROPHYSIOLOGY IN REALISTIC CONTEXTS: BIG HYPE OR BIG LEAP?

Giovanni M. Di Liberto & Emily Y.J. Ip[1,2]

[1] School of Computer Science and Statistics, University of Dublin, Trinity College, Ireland; ADAPT Centre
[2] Trinity College Institute of Neuroscience, Trinity College Dublin, Dublin, Ireland.

**Email contact:** gdiliber@tcd.ie

**Competing Interest Statement:** Disclose any competing interests here.

**Acknowledgements:** This research was supported by Research Ireland at ADAPT, the Research Ireland Centre for AI-Driven Digital Content Technology at Trinity College Dublin and University College Dublin [13/RC/2106_P2]. This work was also conducted with the financial support of the Research Ireland Centre for Research Training in Digitally-Enhanced Reality (d-real) under Grant No. 18/CRT/6224. For the purpose of Open Access, the author has applied a CC BY public copyright licence to any Author Accepted Manuscript version arising from this submission.

## ABSTRACT

Understanding the neural basis of speech communication is essential for uncovering how sounds are translated into meaning, how that changes with development, ageing, and speech-related deficits, as well as contributing to brain-computer interfaces research. While traditional neurophysiological studies have relied on simplified, controlled paradigms, recent advances have shifted the field toward more ecologically-valid approaches. Here, we examine the impact of continuous speech research and discuss the potential of speech interaction neurophysiology. We present a discussion on how realistic paradigms challenge conventional methods, offering richer insights into neural encoding, functional brain mapping, and neural entrainment. At the same time, they introduce significant analytical and technical complexities, particularly when incorporating social interaction. We discuss the evolving landscape of experimental designs – from discrete to continuous stimuli and from socially-isolated listening to dynamic, multi-agent communication. By synthesising findings across studies, we highlight how naturalistic speech paradigms contribute to refining theories of language processing and open new avenues for research. In doing so, this review critically evaluates of whether the move toward realism in speech neurophysiology represents a technological trend or a transformative leap in understanding the neural underpinnings of speech communication.

## ABBREVIATIONS

TRF: temporal response function
EEG: electroencephalography
MEG: magnetoencephalography
OPM: optically pumped magnetometers
ECoG: electrocorticography
fMRI: functional magnetic resonance imaging
fNIRS: functional near infrared spectroscopy
BCI: brain-computer interface
MMN: mismatch negativity
PAC: phase-amplitude coupling
ERP: event-related potentials
ERF: event-related field
AI: Artificial Intelligence
LLM: Large language models

# 1. INTRODUCTION

Speech communication is a fundamental aspect of the human experience, enabling us to express thoughts and emotions, form connections, and foster social cohesion (Hari et al., 2015). Understanding the neural machinery at the foundation of speech communication has great importance over numerous dimensions, from the study of development, ageing, and speech-related deficits to the more fundamental understanding of how sounds are translated into meaning and vice versa. Such insights are also particularly valuable in application, as they inform groundbreaking developments that have been leading to a new generation of brain-computer interfaces (BCIs) (Hjortkjær et al., 2025; Metzger et al., 2023; Raghavan, Patel, et al., 2024).

The neurobiology of speech has been extensively investigated through lesion studies, behavioural experiments, and measurements of neural activity, whether that be electrical or hemodynamic. Such studies have been made possible by using technologies such as electroencephalography (EEG) (Beres, 2017; Kutas & Federmeier, 2010), magnetoencephalography (MEG) (Salmelin, 2007), functional magnetic resonance imaging (fMRI) (Cusack et al., 2018), and functional near infrared spectroscopy (fNIRS) (Chen et al., 2020). While much of that literature probed neural activity using heavily simplified experimental paradigms (e.g., listening to sequences of syllables), research in the last 10-15 years has propelled the study of speech and language neurophysiology toward more realistic tasks. Typical experiments in such spaces involve speech comprehension tasks, where participants are presented with audiobooks or other pre-recorded audio or audio-visual material. Recent developments are moving this research towards even more realistic tasks involving human interaction, facilitated by increasingly accessible hyperscanning technology (EEG, OPM-MEG, fMRI, and fNIRS) (Hirsch et al., 2021; Lin et al., 2023; Misaki et al., 2021; Montague, 2002; Veneziani et al., 2025; Zhou et al., 2025) and other mobile recording technologies, such as wireless EEG (Goregliad Fjaellingsdal et al., 2020). As we approach this potential methodological shift, it is important to understand if the motivation and future advances in this space are primarily technological, or if we might be at the frontier of a groundbreaking new understanding of brain functions.

As is often the case, the reality likely lies somewhere in the middle. There is certainly a technological drive behind these innovations, and there is also potential for new scientific discoveries. The combination of these two exciting prospects can be particularly attractive for scientists. Nonetheless, while these technological developments have already begun to take shape (e.g., hyperscanning technology, mobile EEG technologies), the research vision for exploiting them in speech neurophysiology remains scattered and often blurry. This manuscript attempts to provide some clarity on this line of work and its future directions by discussing two key questions: Have realistic experiments involving continuous speech listening contributed to our understanding of speech neurophysiology? Could the study of more interactive tasks and realistic scenarios contribute to speech neurophysiology with truly novel insights?

After a brief introduction to common experimental designs (**Section 2**), this manuscript discusses those two questions in dedicated sections (**Sections 3 and 4**), concluding with a perspective on key theoretical questions that have the potential of bringing a considerable shift to our understanding of the neural functions underpinning speech processing.

## 2. COMMON METHODS AND DESIGNS

Analysis methods and experimental designs typically come as a pair, as experiments are designed with particular analytical methods in mind. Some designs in speech neurophysiology are rigid, in that they are only suited for particular analyses. Others are quite flexible, in that they can be re-analysed in many ways. This section provides a brief description of some key designs and analytical methods, offering a simple overview of a complex experimental landscape. This overview focusses on receptive speech and serves as a foundation for the subsequent discussion on more realistic experiments.

### 2.1. The discrete-to-continuous dimension

Receptive speech involves the processing of a myriad of acoustic and linguistic properties by a complex and distributed neural network. Real-life speech streams typically involve connected word sequences, with their processing being influenced by their proximal context (e.g., perceptual restoration effect) (Leonard et al., 2016). An additional complexity is that speech units at different timescales are processed simultaneously and in synergy with each other (Gwilliams et al., 2024a). To study particular aspects of that process, tailored experiments can be designed that isolate the property of interest from other possible co-variates. For example, designs that involve listening to isolated syllables or words can be used to pinpoint cortical areas that are implicated in syllable or word encoding and discrimination (Eichele et al., 2005; Shtyrov et al., 2000), while excluding the impact of other linguistic phenomena (e.g., semantics, syntax) by construction. Methodologies such as event-related potentials or event-related fields (ERP/ERF) are common approaches for measuring the average neural activations corresponding to (what is assumed to be) **discrete** events of interest. These experiments can also be designed to study contextual effects by measuring ERP/ERFs in correspondence to discrete speech units occurring in context, as has been done extensively in the N400/P600 literature (Kutas & Federmeier, 2010; Osterhout & Mobley, 1995; van Herten et al., 2005). That literature sheds light on the important synergies between bottom-up and top-down processes in speech processing and on speech development in the first months of life (Kuhl, 2004; Zhao & Kuhl, 2022).

Experiments involving such discrete analytical methods span from technically simplistic designs (e.g., syllable listening) to more challenging ones (e.g., observing syntactic and semantic violations), where the linguistic content must be carefully manipulated to maximise and balance the effects of interest (Yamada & Neville, 2007). One critical downside of this type of experiment is that its tasks may be far from real-world activities, raising questions as to whether our brains operate in the same way "in the wild". This is far from trivial. Teasing apart a process from its covariates can be crucial, but it can also lead to unusual scenarios that our brain might tackle in extraordinary ways.

How can we study speech neurophysiology with more realistic tasks? From a technical perspective, a simple experimental design involves the listening of pre-recorded speech streams from, for example, audiobooks or podcasts. Those speech streams typically consist of **continuous** and coherent sequence of semantic concepts across multiple consecutive sentences. In its basic form, the experimental design simply requires audio recordings, a device for measuring neural activity (e.g., EEG, MEG), and an accurate temporal synchronisation between the two. On the other hand, methods for analysing the multifaceted nature of those data come with some challenges and have only become established and (somewhat) standardised in recent years (Crosse et al., 2016; Crosse et al., 2021a; De Cheveigné et al., 2018; Di Liberto et al., 2024; Di Liberto et al., 2021; Heilbron et al., 2022; King et al., 2020).

The concept of discrete and continuous becomes blurry when considering natural speech. Acoustic properties like the sound envelope are continuous signals, and the study of sound envelope-brain relationships have been extensively explored using methods like the temporal response function (TRF) (Crosse et al., 2016; Crosse et al., 2021b), phase-locking / phase-tracking measurements (Luo & Poeppel, 2007), cross-correlation (Millman et al., 2015), canonical-correlation analysis (De Cheveigné et al., 2018), back-to-back regression (King et al., 2020), and linear mixed effects (Mai et al., 2024). Linguistic units like phonemes and words can also be studied with those methods, which consider such units as part of a time continuum **(Figure 1)**. In these instances, the proximal context can impact the processing of subsequent units, with a likely temporal overlap in their processing. Depending on the specific hypotheses and assumptions, it is also possible to ignore those temporal dependencies and perform analyses (e.g., ERP/ERF) that consider them as discrete isolated events, albeit with some limitations (Khalighinejad et al., 2017).

Experiments involving speech streams can vary greatly in complexity. One of the main difficulties lies in choosing a stimulus, wherein the possibilities of design dimensions are seemingly endless. Such design considerations include number of speakers, scripted versus spontaneous speech, monologue versus dialogue, emotion type and intensity, accent, language, audio quality, and the presence of dysfluencies, among many others. The large number of dimensions highlights the importance of being cautious when drawing conclusions from a single study and of considering appropriate controls and variations. One very positive trend is the rising interest in sharing anonymised data after publication of the results, as this allows scientist to more easily determine whether their results replicate across different datasets, or how they are impacted by factors such as neural recording technology (e.g., EEG vs. MEG), stimulus material, task, and participant cohort.

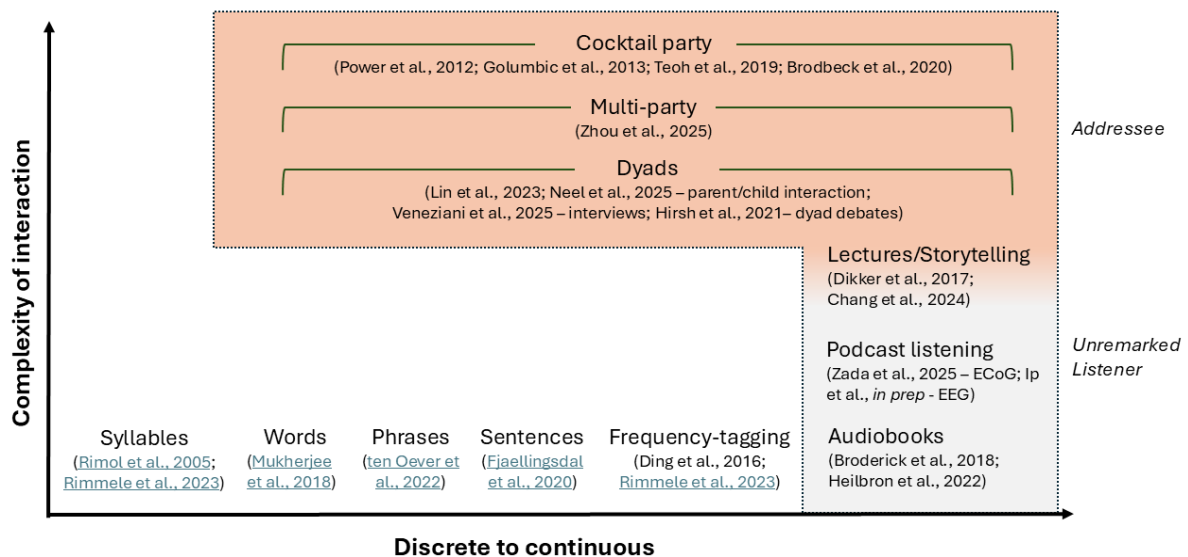

*Figure 1. Diagram capturing typical designs in speech neurophysiology* with a few representative studies for each design. The x-axis represents the type of speech stimulus in the experiment, ranging from isolated syllables and words to continuous speech. Note that the frequency-tagging paradigm is continuous, but not representative of a typical speech interaction. As such, it is placed one step before paradigms involving audiobooks and other natural speech stimuli. The y-axis indicates the transition from experiments only involving receptive speech without any form of interaction to interactive experiments involving a group of two or more people (dyads and multi-party respectively), as well as including scenarios involving many distinct groups of speakers (cocktail party). Note that these more interactive scenarios may also involve tasks (e.g., gamified) where individuals communicate through single words or phrases, as opposed to continuous speech.

Another clever method has also been proposed, namely the frequency-tagging paradigm, which offers an analysis that is somewhat in-between discrete and continuous. This design involves speech stimuli consisting of continuous sequences of words, played in a precise rhythmic manner (Ding et al., 2016; Luo & Ding, 2024). While the artificial nature of that design makes the listening experience quite unusual (but note the potential link with sung speech, which is also rhythmic), the paradigm associates linguistic structures at different timescales with precise and distinct frequencies (Martorell et al., 2023), which can be elegantly studied with a Fourier Transform analysis. Nonetheless, that also leads to a somewhat unnatural experiment, with neural activations that may differ from what is measured with more realistic tasks (Martorell Serra, 2024).

### 2.2. The social dimension

Continuous speech listening experiments have proven effective for studying the multifaceted speech processing network by enabling the measurement of neural signatures of multiple acoustic and linguistic properties with a single experiment. Two key issues should be noted: First, typical experimental *scenarios* involve non-realistic laboratory settings, where participants minimise motor movements and are alone in a dark sound-proof room (Levy et al., 2025); Second, while these tasks are relevant to the real-world, they do not capture the most important aspect of speech: Social interaction. While controlled experiments afford the ability to isolate specific properties of interest, the main challenge is designing experimental set-ups that allow researchers to disentangle the concurrent phenomena that complicate interaction, both exogenous (e.g., background noise) and endogenous (e.g., trust, social anxiety, empathy), while allowing participants to communicate freely as if interacting "in the wild."

Designing speech interaction experiments is possible and has been explored in specific subfields. These experiments enable the study of a whole range of phenomena that are particular or especially important for social interaction, such as turn-taking (Bögels, 2020), brain-to-brain synchronisation (Pérez et al., 2017a), and neural encoding alignment (Zada et al., 2024). However, this opportunity for studying speech neurophysiology comes at a cost: the social dimension adds an enormous number of factors, as well as non-negligible technical complexities.

What could be an ideal experimental design to study the neurophysiology of speech interaction? Attempts at outlining various design elements in multi-speaker scenarios have reached limited success, as considerations in experimental design span across norms of diverse disciplines and modalities (Gregori et al., 2023; Offrede et al., 2021). For example, one could record neural data from participant dyads as they freely interact. Such paradigms could illuminate how brains in dyads process speech during listening (from sound to meaning), tease apart the neural processes enabling speech production, quantify brain-to-brain synchronisation, and derive neural metrics for studying the many other factors that have a role in speech communication. Of course, a design of this kind comes with many complexities. For example, challenges arise regarding how to prompt the conversation, how to ensure that the dialogue goes both ways, how to control for individual biases (e.g., social biases), social anxiety, and other phenomena. Furthermore, the number of agents and their roles are key factors that can lead to different types of social interaction. For instance, consider that an agent might be human or based on artificial intelligence (AI) (Diederich et al., 2022); an agent may play different roles in an interaction (Bell, 1984); a meeting might be in person, virtual, or hybrid; an interaction might involve two or more parties (Branigan, 2006); the speech might be primarily unidirectional (e.g., one-to-many lecture) or bidirectional (e.g., conversation); the social context (e.g., job interview vs. casual conversation). All these scenarios pose various degrees of speech accommodation and audience design factors and must be taken into account during neurophysiological experimental design.

One factor of particular importance is how an agent may play different roles in an interaction. Different language has been used to refer to these audience role distinctions, where early work established the terminology "speaker, addressee, auditor, overhearer, and eavesdropper" (Bell, 1984). More recently, the terminology has evolved as such concepts have made its way through other disciplines, including the terms "second-/third-person neuroscience" in the last few years (Li & Zhang, 2024; Redcay & Schilbach, 2019). Our discussion surrounding this topic will introduce the term of an "unremarked listener" **(Figure 1)** role to refer to the previously used term "eavesdropper," as we give an overview of the research towards social speech neurophysiology. We later highlight the importance of harmonising terminology across related fields in our final reflections.

There are also purely technical challenges, for example, related to the simultaneous recording of neural data from multiple individuals at once, or hyperscanning. This technology has been explored for at least 20-25 years (Flexer & Makeig, 2007; May, 2004; Montague, 2002), and recent developments facilitate technical challenges of data collection, such as data synchronisation. Other challenges involve the drop-out rate when testing dyads (one participants dropping out typically means cancelling the session for both participants), as well as the transcription and alignment of the recorded speech for the purpose of feature extraction and analysis. All of these challenges must be considered, not to mention other very important dimensions, such as the visual modality (e.g., non-verbal cues).

While designs involving such a multitude of factors and mechanisms have been explored (Goldstein et al., 2024; Xie et al., 2023), most studies in the literature have controlled for some of those complexities by simplifying either the analysis or the experimental paradigm. For example, turn-taking studies typically constrain their analyses to the measurement of ERP/ERFs surrounding the speaker-switching moments, with a design that can address realistic speech interactions (Bögels, 2020; Kuhlen & Abdel Rahman, 2023). Another relatively simple analysis that can be applied to realistic interaction designs is the study of brain-to-brain synchronisation (Bevilacqua et al., 2019; Dikker et al., 2017; Montague, 2002; Pérez et al., 2017c), which can be used to pinpoint similarities between individuals, but is less suited for disentangling the neural processing of different linguistic properties. Other studies opted for design simplifications, for example involving the recording of only one participant in a dyad during the speech interaction (Goldstein et al., 2025). Another simplification involves scenarios where participants are presented with the same stimulus simultaneously while in the same room, without speaking to each other (Dikker et al., 2017).

Note that, while our discussion focusses on the two continua above, there exist other important dimensions of interest that we do not cover in this manuscript. One important dimension is the visual modality, including verbal and non-verbal cues that have a central role in social interaction (Gregori et al., 2023). Another important area of study is the speech production side of the interaction, which we don't cover explicitly in this manuscript.

## 3. DISCOVERIES WITH DESIGNS INVOLVING SPEECH STREAMS

Early neurophysiology research studying the listening of continuous speech explored phenomena such as stimulus-neural synchronisation and neural-neural synchronisation (e.g., correlation between repetitions, cortical rhythms, or individuals) (Luo & Poeppel, 2007; Peelle & Davis, 2012). When recording neural signals during speech listening, the neural traces naturally reflect a myriad of processes, spanning from sound to semantics. One key challenge is to determine what exactly contributes to the neural signal, characterising temporal, spatial, and spectral origins of its components. One approach to isolating a neural process of interest involves contrasting neural activations in tailored experimental conditions. For example, contrasting neural responses to intelligible and unintelligible speech has elucidated what cortical sites are involved in speech comprehension, besides sound processing (Di Liberto et al., 2015). More recently, multivariate regression methods were proven effective for analysing neural signals recorded during speech listening, enabling the isolation of speech processing stages from a single neural trace, without the need for tailored conditions (Brodbeck et al., 2018; Broderick et al., 2018; Di Liberto et al., 2021; Di Liberto et al., 2015). Advantages include the possibility of measuring multiple neural metrics with a single experiment, the use of pleasant listening tasks, and the reduced reliance on artificial conditions (e.g., listening to nonsense speech).

This section discusses some of the key impacts of speech stream neurophysiology. While we do not intend this section to be comprehensive, we mention some of the most distinctive research contributions to date, aiming to spark reflections on how the study of speech streams has contributed to our understanding of speech neurophysiology. Rather than detailing a chronological list of findings, we focus on key areas of interest in this space: neural encoding of linguistic units, functional brain mapping, neural entrainment, and decoding **(Figure 2)**.

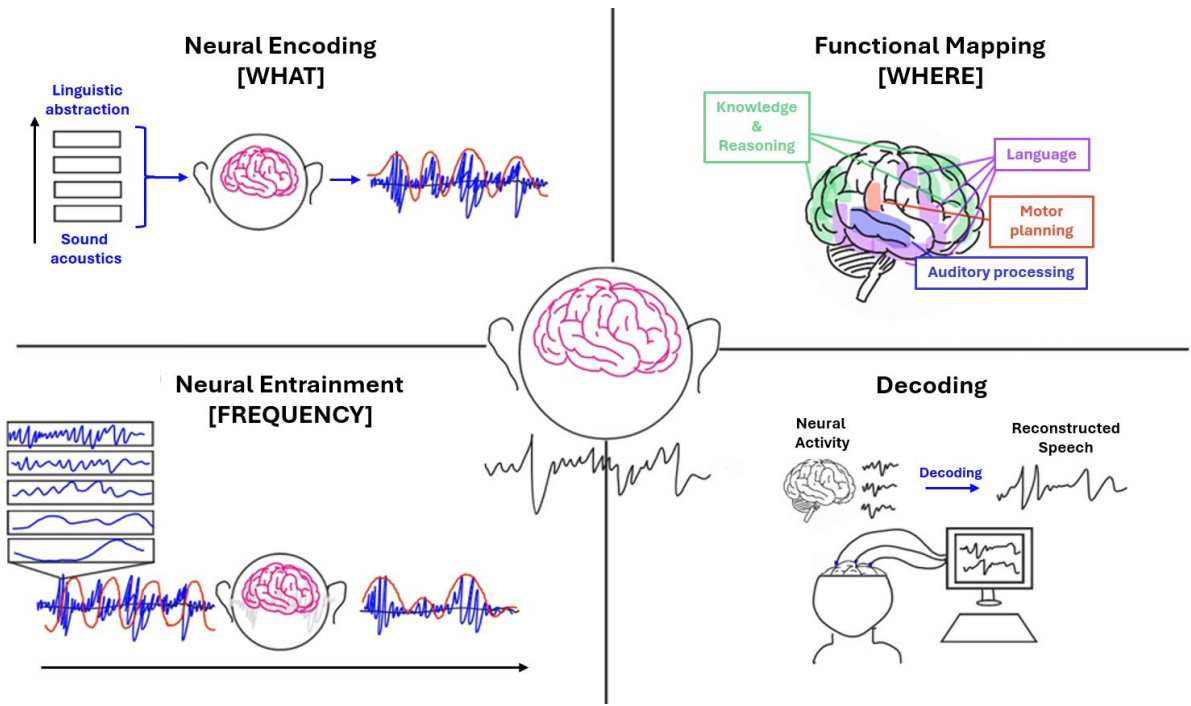

***Figure 2. Key areas of interest in speech neurophysiology****. Research contributions in speech neurophysiology typically focus on determining what information our brain encodes, where that is encoded in the brain spatially, and what neural rates are involved. Decoding research may combine these various dimensions with objectives in areas such as brain-computer interfaces and objective markers. Each of these four areas is described in a dedicated subsection.*

One challenge is that studies on speech stream processing can vary in many aspects, such as experimental paradigms and participants cohorts, as well as involving the study of neural dimensions. Results from such experiments may shed light on what language properties are encoded in the neural signals, where they are encoded, how their encoding unfolds over time, and what neural rhythms are most sensitive to a given speech or language property. For example, let's consider studies relating developmental dyslexia with atypical EEG delta-band (~0.5-4Hz) stimulus-neural synchronisation (Power et al., 2016), or a study relating that same frequency band with speech comprehension (G. M. Di Liberto et al., 2018). Taken in isolation, those studies are useful characterisations of the role of delta-band neural activity. But how is that valuable exactly? Reviewing that work can help connect results from different studies, providing added value such as the formulation or validation of theories relating to speech processing.

### 3.1. Neural encoding of linguistic units

How does the human cortex transform speech into abstract meaning? What speech and language properties are explicitly encoded in our brain? How does that encoding change over the course of our lives, for example throughout childhood? These and other related questions have been extensively studied via behavioural and neural investigations, generating a large repertoire of results and numerous theories. While many unanswered questions and ongoing debates remain, studying the neurophysiology of speech stream processing has greatly contributed to that research.

Multivariate analysis methods have proven effective in measuring the neural processing of a multitude of acoustic and linguistic properties, when applied to neural data recorded during continuous speech listening. While early research was primarily concerned with the neural encoding of the sound envelope (Lalor & Foxe, 2010), numerous studies in the last decade or so have shed light on the encoding of linguistic properties at the phonological (Di Liberto et al., 2015), syntactic (Brennan & Hale, 2019), prosodic (Gnanateja et al., 2025; Teoh et al., 2019), and lexical levels (Brodbeck et al., 2018; Broderick et al., 2018; Chalehchaleh et al., 2025). That work primarily involved listening tasks using continuous speech recordings, for example, from audiobooks and podcasts. The possibility of measuring the neural processing of those properties with realistic listening tasks and from a single EEG trace is attractive, and it is also valuable in that it enables re-analyses of existing datasets from different angles. Other clever designs were also proposed involving more controlled stimulation, such as the frequency-tagging paradigm (Ding et al., 2016), where words are presented in a rhythmic manner, with different timescales of the linguistic structure being associated with distinct frequencies (for example, 4Hz for syllables, 2Hz for words, 1Hz for phrases).

The study of speech streams has provided a rich set of (often) confirmatory evidence on the neural encoding of language. One main contribution is the generalisation of previous findings to more realistic listening tasks, increasing the level of detail and clarity on the encoding of distinct stimulus properties. For example, the temporal progression of the speech to meaning transformation proposed based on ERP/ERF results (Salmelin, 2007) could not clearly resolve the temporal overlap of the various responses to sound, phonology, and lexical-semantic analysis. However, the study of continuous speech streams involves regression methods designed to tackle that temporal overlap, carrying out that separation across different timescales (Di Liberto et al., 2021; Di Liberto et al., 2015), informing us about how our brain handles that temporal overlap within a given timescale (Gwilliams et al., 2022). Such insights may even lead to theories that specifically address the retention of information over time in presence of that temporal overlap (Gwilliams et al., 2024b).

In the context of phonological processing, there is an ongoing debate whether phonological abstractions are relevant to how our brains process speech streams (Daube et al., 2019; Port & Leary, 2005; Ramscar & Port, 2016; Stilp & Kluender, 2010). Previous research using audiobook listening tasks suggested that neural responses to continuous speech reflect the acoustically-invariant encoding of phonemic identities (Di Liberto et al., 2015). That work used careful control analyses to isolate neural correlates of acoustically-invariant phonological information. Since accounting for all possible acoustic features is not possible, that result initially raised justified critiques (Daube et al., 2019). Further research has been conducted to address that concern, presenting results that support the initial hypothesis by relating those phonological measurements with phonological awareness (Giovanni M. Di Liberto, Edmund C. Lalor, & Rebecca E. Millman, 2018; Giovanni M. Di Liberto, Varghese Peter, et al., 2018), comprehension (G. M. Di Liberto et al., 2018; Millman et al., 2015), language proficiency (Di Liberto et al., NeuroImage, 2021), early development (Di Liberto et al., 2023), hearing impairment (Carta et al., 2024), speech register (Piazza et al., 2025) and across multiple languages (Carta et al., 2024; Lesenfants et al., 2019). That work was also recently validated and extended by a team using intracranial data (Mai et al., 2024) and has been complemented by the finding that phonological processing persists with a substantial overlap between neighbouring units (Gwilliams et al., 2022) and is influenced by predictive mechanisms (Brodbeck et al., 2018; Di Liberto et al., 2023; Di Liberto et al., 2019). Overall, these results relying on speech streams support the view that phonological processing is central to speech processing in the human brain, with rich insights into the cortical regions involved, such as the superior and lateral temporal cortex, ventral/dorsal parietal cortex, and inferior prefrontal cortex (Gong et al., 2023; Mesgarani et al., 2014). These results have far-reaching consequences for phonological theory and theories of language representation and learning in the brain (Di Liberto et al., 2023).

The study of speech streams has also impacted our understanding of early speech development. Previous work with sound discrimination tasks have already offered compelling evidence that infants specialise their speech processing to the language they are learning between 6 and 12 months of age (Kuhl, 2004). However, the focus on basic phonetic contrasts, often using isolated sounds, did not necessarily tell us how the human brain encodes that distinction, or how that phenomenon is relevant to continuous speech processing. Studies on the neurophysiology of speech involving continuous streams offer that complementary view, confirming and refining previous results by informing us on the neural distinctions in the encoding of different speech sounds. While a large part of that research focusses on neural synchronisation metrics (see **section 3.3**), there is also evidence that an acoustically-invariant encoding of phonetic features starts forming between 4 and 7 months of age (Attaheri et al., 2022). This finding generalises previous research based on sound discrimination, indicating a time-window where the distinct encoding of phonetic feature is sufficiently consistent over time when presented with continuous speech streams (in that specific case, in the form of sung or spoken nursery rhymes).

Similar research has been carried out that studies the neural processing of other linguistic constructs, shedding light on how, where, and when they are processed when listening to continuous streams. Syntactic structure has been associated with various cortical sites, with research on continuous speech streams playing a key role in the anterior temporal lobe (Brennan et al., 2012; Brennan et al., 2016). While that association was initially found using manually labelled predicate argument structures, more recent evidence examined how well computational models of parsing effort capture neural activity during speech listening. For example, combinatory categorial grammars captures have been shown to be a better model of the neural processing of language structure than context-free grammars and large language models (LLMs) (He et al., 2024; Stanojević et al., 2023). Other research has explored the neural encoding of prosody, finding the Heschl's gyrus, the superior temporal plane, and the superior temporal gyrus to be central to its processing (Gnanateja et al., 2025), with the delta-band being of particular importance (Teoh et al., 2019).

Another important property of speech streams involves the temporal regularities of sound and language. These regularities are present at various timescales, for example concerning phoneme sequences (phonotactics), syntactic units, word form, and semantic meaning (Kaufeld et al., 2020; Keitel et al., 2018). The neural processing of those regularities has been studied in-depth for decades, forming an extensive literature on semantic and syntactic expectedness and violation (Bornkessel-Schlesewsky & Schlesewsky, 2008; Kutas & Federmeier, 2010; Kutas et al., 1984/01; van Herten et al., 2005). While that research typically focussed on expectations at a single linguistic level, studying the processing of speech streams can inform us about the simultaneous unfolding of neural predictions at those multiple linguistic levels. In contrast with the "reactive" perspective, where the focus is the neural consequence of an unexpected event, other research has also found that linguistic predictions shape the early encoding of speech sounds (Broderick et al., 2019; Dou et al., 2024) and even reported anticipatory neural activations reflecting those predictions (Goldstein et al., 2022; Leonard et al., 2016). Once again, studying speech streams presents favourable conditions for studying these phenomena from a single neural trace.

### 3.2. Functional brain mapping

The previous section concerns what properties are encoded in the human cortex during speech listening and what processes are involved (e.g., lexical prediction). Another important angle involves the precise identification of the spatial origins and pathways enabling speech listening. Models like the dual-stream model have guided neurophysiology research in this area for about two decades (Hickok & Poeppel, 2007). The study of speech streams combined with advances in neural recording technology has led to a more detailed brain mapping of the neurophysiology of speech listening. Previous work often involved assumptions involving, for example, the sequential nature of the pathways transforming speech into meaning. Indeed, that simplification is often carried out for practical reasons, as there is convincing evidence on a complex interplay of bottom-up and top-down processes, as well as results pointing to parallel pathways even at the acoustical-phonological interface (Hamilton et al., 2021).

The resolution has also increased (Chung et al., 2022). Evidence from fMRI and intracranial EEG substantially magnified our previous understanding on the spatial organisation of speech in the human cortex (Cusack et al., 2018; Formisano et al., 2008). More recently, findings from intracranial recordings detailing the neural encoding of speech sounds in the human auditory cortex (Mesgarani et al., 2014) have been revisited by evidence from higher-density intracranial data, even at the level of single-neurons (e.g., Neuropixels arrays) (Leonard et al., 2024). Those results are partly confirmatory, as they found neural populations across the depth of cortex to be tuned to a dominant speech feature, consistent with previous intracranial electrocorticography (ECoG) results (Mesgarani et al., 2014; O'Sullivan et al., 2019). On the other hand, neurons also show encoding to other speech features across the depth dimension, indicating a more complex picture than previously observable that highlights the complexity and varied neural encoding across the cortical layers (Baratham et al., 2022). This brain mapping will likely lead to a rich set of data in the coming years, sparking a reflection: How are we going to harness this high-density data? Could those data answer some of the ongoing debates, or is this a different dimension opening a whole new set of questions?

Recent developments in deep learning technologies have introduced a further dimension to this research. Indeed, our field is interested in mapping brain areas to their function, and to identify the corresponding neural pathways. In the context of speech, the longstanding approach has been to relate known speech and language features of interest with the neural data in different sites. Those features were typically derived from transformations or models from linguistics, that we either fully understand or that have been tested in various domains e.g., phonological identity and alignments (Di Liberto et al., 2015); phonotactics and syntax from rule-based approaches (Brennan & Hale, 2019; Brennan et al., 2016; Di Liberto et al., 2019). LLMs opened up the new possibility of providing reliable features at multiple levels of complexity that are not derived in an explicit manner. For example, decoder-only transformers, like the well-known LLM GPT-3, are trained using a self-supervised approach on next-word prediction. In other words, they are not trained to explicitly encode phonotactic rules, syntax, or anything in particular. Since they are based on text, it is not obvious that they would be a good model for the neural processing of speech. Nonetheless, numerous studies have successfully associated LLM-derived features to neural signals (Chalehchaleh et al., 2025; Heilbron et al., 2022; Tikochinski et al., 2023). While some of those studies focussed on specific levels of processing (e.g., next-word prediction), other studies related progressively deeper hidden layers of open-source LLMs with neural signals, producing spatio-temporal maps and enabling the data-driven identification of neural pathways for the speech-to-meaning transformation (Caucheteux et al., 2022, 2023; Mischler et al., 2024).

Does this mean that rule-based or linguistic-driven features are not useful anymore? Not necessarily. These models are certainly interesting and useful for characterising the speech processing hierarchy altogether. Potential limitations such as their text-based nature may be overcome by ongoing developments in speech transformers. Nonetheless, there are several issues that should be considered. Speech and language features derived from LLMs or other similar models are not guaranteed to be interpretable or to match brain functioning. In fact, it is quite likely that the processing is similar to brain functioning in some respects but not others. LLM development is not bound to such concepts, and it is possible that future developments will lead larger models away from neurophysiological relevance, if that's convenient for the model objective (e.g., next-token prediction). So, one quest might involve identifying a set of most neurophysiologically plausible LLMs to be used in our field. That would at least bring some consistency to an area of research whereby different models are used across studies, even within the same research team. Appropriate data, code, and model sharing becomes crucial in this context (Di Liberto et al., 2024; Makeig & Robbins, 2024) as results using new LLM models will have to also be compared with other LLM models used in previous studies. Failure in doing do carries a risk for *p*-hacking. For example, let's say that we obtain a nice-looking result with a given LLM after failing with other five LLMs. Failing to acknowledge the negative results when publishing the study would constitute a big problem, as that would lead to a literature with a diverse set of models that succeed, while in reality those models might have failed in other studies. This is a difficult challenge that can be tackled via appropriate data sharing, as other researchers may then evaluate the consistent neurophysiological validity of an LLM across tasks, cohorts, and recording modalities. While that issue also exists for linguistic-driven models, LLMs increased the risk substantially due to the large number of possible models that one might select. In sum, our view is that this is a powerful new tool, where the excitement is matched by the risks. So, the future will probably involve a combination of these data-driven methods with linguistic-driven hypotheses and controls. One recommendation that we put forward in this context is to have strong motivations in place before using a certain model (e.g., a certain version of an LLM). For example, a previous study may have used that model, or a particular fine-tuning was applied that might be relevant to the particular research question at hand. The other recommendation is indeed to report the negative results as well, as they might be important, especially when interpreting a result in the context of the literature or when aggregating results from different studies and investigators.

### 3.3. Low-frequency neural entrainment

Neural signals at different rates align with incoming sounds. This phenomenon, known as neural entrainment in the broad sense, or neural tracking, has been associated with various potential underlying mechanisms, such as evoked responses (Oganian et al., 2023), endogenous neural oscillations synchronising to the stimulus (Obleser & Kayser, 2019), and other endogenous activations related to auditory predictions and sensory restoration (Brodbeck et al., 2020; Cervantes Constantino & Simon, 2018; Leonard et al., 2016; Morillon & Schroeder, 2015). Despite being a complex phenomenon with ongoing debates about its contributors, as in (Chalas et al., 2023; Mai & Wang, 2023) and (Duecker et al., 2024; Haegens, 2020; Lalor & Nidiffer, 2025), researchers have extensively studied its relationship with factors such as comprehension and attention (Broderick et al., 2018; Heilbron et al., 2022; Power et al., 2012), in addition to investigating how it is impacted by brain development, ageing, hearing impairment, neurodiversity, and speech-related deficits (Broderick et al., 2021; Carta et al., 2024; Giovanni M. Di Liberto, Varghese Peter, et al., 2018; Kries et al., 2024; Przysinda et al., 2025; Wynn Camille et al., 2018). The use of speech streams for these investigations has substantially advanced our understanding of this phenomenon, informing us about the potential functional roles of different neural rates for speech perception.

The exact role of different neural rates in speech processing remains under debate, with a large set of results that are at times difficult to reconcile. Delta (~0.5-4 Hz) and theta (~4-8 Hz) band rates are crucial for neural tracking of the acoustic envelope, potentially aiding speech segmentation and syntactic processing of linguistic units at timescales corresponding to their cycles (Martorell et al., 2023; Peelle et al., 2013). Other studies have shown that brain-acoustic entrainment in delta and theta co-varies with speech intelligibility (Doelling et al., 2014; Mai et al., 2016; Peelle et al., 2013), with delta linked to comprehension and theta to speech clarity (G. M. Di Liberto et al., 2018; Giovanni M. Di Liberto, Edmund C. Lalor, & Rebecca E. Millman, 2018; Etard & Reichenbach, 2019). However, other findings suggest that delta is more related to speech chunking and discourse-level parsing than meaning (Boucher et al., 2019) and theta is entrained more generally to the processing of acoustic attributes (Boucher et al., 2019; Kösem & van Wassenhove, 2017). In fact, delta-rates appear to have an important role since birth, with some proposing that they might be central for learning speech segmentation, even before the processing of word meaning (Attaheri et al., 2022). From there, the alpha band (~8-13 Hz) is a hallmark of the higher-level sensory and cognitive processes involved in speech- and language-related processing and comprehension, such as working memory and attention. This has been demonstrated in active listening tasks measuring speech intelligibility (Dimitrijevic et al., 2017) and comprehending speech under adverse conditions (Obleser & Weisz, 2012). Such processes of attention are modulated in EEG responses in studies that investigate selective attention in speech (Näätänen et al., 2002; Power et al., 2012). Other studies highlight the function of alpha rates, alongside beta rates, in other low-level mechanistic operations facilitating language comprehension (Zioga et al., 2023), as well as the unification of semantic and syntactic information in a "desynchronization effect" (Lam et al., 2016).

A similar heterogeneity can be observed when considering higher neural rates. Beta (~13-30 Hz) rates were reported to be involved in syntactic processing (Bastiaansen et al., 2010), though predictive mechanisms implicate them in the semantic processing of upcoming words when coupled with gamma rates (Meyer, 2018). Early studies have suggested gamma rates (30-70 Hz) to also be involved in lexico-semantic retrieval in EEG/MEG (Lutzenberger et al., 1994; Pulvermüller et al., 1997), while a wide range of literature suggests neural signals at gamma rates to be related with sound amplitude modulations at subsyllabic frequencies (Lizarazu et al., 2019; Meyer, 2018). Predictive processing has become increasingly adopted in recent years to frame functions of oscillations, whereby top-down predictions in the beta rates mediate sentence-level representation and facilitate comprehension (Lewis et al., 2016; Mamashli et al., 2019). Moreover, gamma power was found to be correlated with an upcoming word's predictability rather than semantic integration (Lam et al., 2016; Wang et al., 2012), further supported by findings that high-gamma frequency neural signals index bottom-up propagation of the prediction error (Prystauka & Lewis, 2019). This suggests complementary functions at the beta and gamma rates in the predictive processing framework (Meyer, 2018; Weissbart & Martin, 2024), while the functions of these bands in the frequency domain are posed as distinctly segregated (Bastiaansen & Hagoort, 2015).

A wealth of research has also been conducted on the role of phase-amplitude cross-frequency coupling (PAC), and has shed light on many facets implicated in speech neurophysiology such as speech processing and behaviour (Arnal et al., 2015; Attaheri et al., 2022; Lakatos et al., 2008; Lizarazu et al., 2019; Mai et al., 2016; Meyer, 2018), interaction between memory systems (Fell & Axmacher, 2011; Lisman & Jensen, 2013), and oscillation-mediated predictive processing mechanisms (Weissbart & Martin, 2024; Zoefel & VanRullen, 2015), among others. This has yielded key insights in children's acquisition of speech (Goswami, 2019), dynamics of cross-language communication (Liu et al., 2021), and characterising speech discrimination behaviour of hearing-impaired individuals (Wang et al., 2023), to name a few.

In addition, research has been substantially augmented by studying neural entrainment in participants cohorts with particular deficits or neuro-diversities. For example, impaired neural entrainment in delta-rates has been associated with difficulties in crucial speech processing tasks, such phonological awareness in developmental dyslexia (Giovanni M. Di Liberto, Varghese Peter, et al., 2018; Goswami, 2011). A similar parallel is drawn in neurodiverse populations, such as in children and adults with spectrum disorder who experience impaired speech rate entrainment (Wynn Camille et al., 2018), and cohorts that experience abnormal neural entrainment (e.g., attention-deficit/hyperactivity disorder) and selective-attention deficits (Calderone et al., 2014; Obleser & Kayser, 2019)

This section only briefly covers part of the literature, supporting one simple reflection: more work is necessary to reach a consensus on what neural entrainment in the broad sense reflects exactly. The large number of dimensions to consider when analysing that literature (e.g., experimental design, stimulus material, analysis method, neural recording modality) may mean that apparent contradiction might be reconciled. In fact, it is possible that entrainment in a given band (or PAC) reflects numerous phenomena, some of which are more prominent than others depending on the experimental paradigm and analysis. This complex landscape of results highlights the extensive effort to study this phenomenon, with heterogeneous methodologies and results that are complementary and contradictory at times, likely influenced by specific stimuli and tasks. Certainly, more work is needed (Chalas et al., 2023).

One angle is that the finding might contribute to the formulation or validation of a theoretical model. For example, the development of the Temporal Sampling Framework of developmental dyslexia (Goswami, 2011) was informed by a variety of both behavioural (e.g., finger tapping tasks) (Thomson & Goswami, 2008) and neural evidence, including studies examining the neural encoding of speech rhythms, and then further tested with tailored experimentation (e.g., (Flanagan et al., 2025)). This led to a comprehensive model that explains how diversities in the processing of different speech rhythms can lead to phonological and reading difficulties. The study of speech streams and the corresponding neural activity was a crucial component to that research, complementing behavioural evidence by offering a more direct view into how speech and neural rhythms align in different cohorts of participants.

### 3.4. Speech decoding and brain-computer interfaces

Speech neurophysiology has also a brain-computer interface dimension. While developments have been initially slow in this area, the last decade or so has seen a growing interest in auditory attention decoding interfaces. Previous research demonstrated that, given a multi-talker scenario, attended and unattended speech streams are encoded differently by our brains (Power et al., 2012; Zion Golumbic et al., 2013). That difference can be detected by neural recording technologies such as EEG, enabling the robust decoding of the focus of attention (Hjortkjær et al., 2025; O'Sullivan et al., 2015; Raghavan, Patel, et al., 2024; Straetmans et al., 2024). The finding is regarded as a breakthrough and has attracted a substantial interest both in academia and industry. Part of that work progressed to investigate the neural underpinnings of attention mechanisms across various cohorts of interest (Carta et al., 2024; Choudhari et al., 2024; O'Sullivan et al., 2019), while other work further investigated attention decoding for brain-computer interface development (Han et al., 2023; Raghavan, O'Sullivan, et al., 2024; Raghavan, Patel, et al., 2024). We look forward to seeing future developments in this direction, which will likely involve personalised multimodal systems combining biosignals of various kinds.

Another very exciting recent development involves transcription of brain signals to text or speech. While this research is primarily possible using invasive recordings, for example, in patients that are fully paralysed (Metzger et al., 2023; Metzger et al., 2022; Moses David et al., 2021; Pei et al., 2011), recent developments have also explored with some success the possibility of decoding text from brain signals recorded with MEG (Défossez et al., 2023) (Lévy et al., 2025) while EEG was much less accurate in that specific study. This recent development might lead to the most naturalistic brain-to-text decoding approach to date, going beyond the well-known P300 speller paradigm (Farwell & Donchin, 1988). The brain-to-text decoding may be performed based on various tasks, from attempted hand-writing movements (Lévy et al., 2025) to imagined speech articulation (Pei et al., 2011). While the majority of these non-invasive decoding studies involve speech listening or text typing recordings, we expect that future studies will investigate proof-of-concept for these methods, for example in paralysed patients attempting to speak or type.

In summary, this section indicates that studying speech neurophysiology using continuous sound streams has led to important scientific advances. Some of the progress involves the generalisation of previous findings to the more realistic context, confirming or complementing prior models and results. This line of work produced new insights into how our brains cope with the serial relationships and high temporal overlap between linguistic units during speech listening. This is novel in that previous research attempted to avoid having those temporal overlaps in their experiments, due to the analytical challenge that they pose. The study of speech streams also enabled the investigation of how speech processing is supported by more general neural mechanisms, such as neural entrainment. Finally, decoding research in this domain has been attracting quite some interest across academia and industry in recent years, bringing us much closer to real-world applications (Metzger et al., 2023).

## 4. SOCIAL SPEECH NEUROPHYSIOLOGY

Having clarified the key advantages of studying the neural processing of speech streams, a natural question arises: Would it be beneficial to explore the neurophysiology of even more realistic scenarios involving social interaction? This section presents early results on the study of social speech and discusses our perspective on which directions might be precursors to scientific breakthroughs.

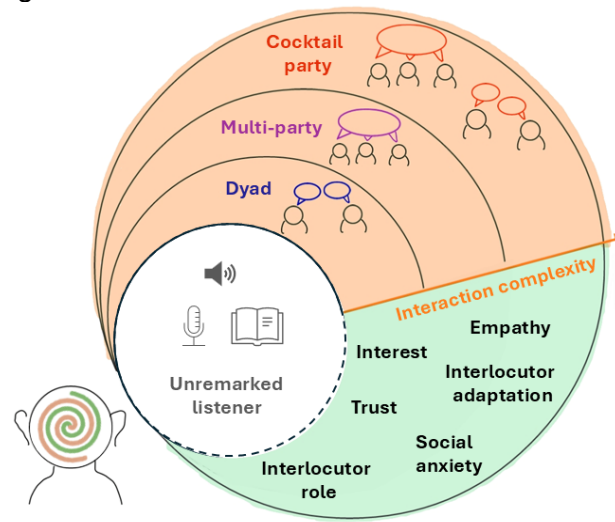

***Figure 3. Opportunities and challenges of social speech neurophysiology.*** *This diagram highlights that social scenarios require considering numerous factors that are typically controlled for when studying receptive speech in social-isolated lab scenarios. The interaction complexity axis (orange) indicates four different levels of interaction, ranging from no interaction, where the speech material does not account for the listener (unremarked listener), to interaction involving one group with two or more interlocutors (dyads and multi-party respectively), or multiple groups (cocktail-party). The green area points to some endogenous processes that are known to have a role in social interaction, and have been extensively studies in other research domains. In other words, these are key factors that should be considered when introducing the social dimension in speech neurophysiology research (green).*

### 4.1. Beyond audiobook listening

What could be more realistic that listening to audiobooks? Audiobook listening is a real-world task, which is also similar to common activities like radio and podcast listening. However, those tasks do not capture the most important use of speech: social interaction.

The social dimension of speech encompasses a wide range of research domains, including anxiety, trust, empathy, communication accommodation, and disorders affecting social interaction **(Figure 3)**. While some of these domains have been studied in controlled scenarios, such experiments may be far removed from the way our brains function during real speech interactions (Hung et al., 2024; Rigoulot et al., 2014). This discrepancy between audiobook listening and interactive speech is akin to the syllable-to-stream gap discussed in the previous section. Investigating speech interaction through neurophysiology could provide fresh insights into ongoing research, as well as opening up new research directions.

Early research on continuous speech streams raised concerns on whether it would be feasible to study neural responses to such a complex process involving the simultaneous processing of many stimulus features. Appropriate methodologies, like multivariate TRFs, were identified for tackling that challenge, without excessively compromising on the naturalness of the listening task. It remains unclear whether those methods would be sufficiently effective with speech interaction tasks, considering the myriad of linguistic and non-linguistic factors at play. The other key concern involves the (probably) inevitable methodological complications that come with the study of speech interaction. Recording data from two or more participants simultaneously increases risks such as participant drop-out, issues with the recording device (e.g., battery, damaged sensors), as well as introducing challenges such as EEG-EEG synchronisation. Modern devices developed for hyperscanning minimise those risks, providing automation and control to such experimental setups (Holmes et al., 2023). While those challenges are common across all hyperscanning experiments, speech interaction adds further complications like sound recordings. Those sounds must be synchronised with the EEG and, depending on the target analysis, transcribed, temporally-aligned, and preprocessed for behaviours that emerge in spontaneous natural speech. For example, dysfluencies are interruptions that occur in the flow of speech, and can happen voluntarily when interlocutors in an interaction utilise them to regulate and guide the course of conversation (Clark & Tree, 2002; Kowal, 2009) or involuntarily in cases like stuttering (Mulligan et al., 2001). We address dysfluencies in further detail in the following section. All these processing steps are technically much more challenging with speech interaction than audiobook stories read by a professional in a quiet environment, raising concerns about the possibility of accurately automating this process, which is crucial for scalability and cost-effectiveness of these experiments. Indeed, the ongoing work has explored experimental or analytical simplifications that would still allow for the study of speech interaction in reasonably realistic setups, which we discuss below.

The technical challenges of studying the neurophysiology of speech interaction have attracted the attention and excitement of some researchers in academia and industry, advancing hyperscanning technology and producing encouraging early results (Hirsch et al., 2021; Holmes et al., 2023; Lin et al., 2023; Misaki et al., 2021; Veneziani et al., 2025; Zhou et al., 2025). Despite important progress, some technical challenges and experimental encumbrance may be hard to avoid; these experiments will likely remain more complex and less accessible than audiobook listening experiments for the time being. Without a doubt, the wide range of possible experimental factors calls for a clear formulation of hypotheses, replication, controls, and larger sample sizes, even more so than when studying continuous speech streams. In other words, the risk for false positives or result misinterpretations will be higher than before if the design is taken too lightly. So, are these experiments feasible, and are they worth it?

Before diving into this, it should be noted that there are other research directions going beyond audiobook listening. One noteworthy idea is the study of speech-in-the-wild, where the neurophysiology of speech listening is studied in out-of-lab settings (simulated or real; (Daeglau et al., 2025; Haupt et al., 2025; Korte et al., 2025; Straetmans et al., 2024)). For example, recent work explored proneness to distraction and mind-wandering using a hybrid approach involving virtual reality and a real classroom space (Levy et al., 2025). This and other lines of work would deserve dedicated reviews and are not further discussed here.

### 4.2. Preliminary impact

Recent research has explored the possibility of probing the neural processing of speech streams in more socially-relevant scenarios. While the literature includes investigations considering a variety of scenarios, from dialogue listening (Ip et al., in preparation; Zada et al., 2024; Zada, Nastase, Aubrey, et al., 2025) to dyadic interactions (Liu et al., 2019; Neel et al., 2025; Uhrig et al., 2018; Veneziani et al., 2025; Zada, Nastase, Speer, et al., 2025), that body of work remains sparse and constrained to the study of specific research questions via targeted controls. However, placing those results in the context of the broader literature is much harder due to the various experimental choices and diverse social components studied. Here, we attempt to provide some clarity on what has been done and what might be possible to study in the future of social speech neurophysiology, organising some of the key findings according to the social dimension (i.e., complexity of interaction) in **Figure 1**.

A first step into the social dimension involves speech listening scenarios with a social component, like **pre-recorded dialogues between two or multiple speakers** (e.g., from podcasts). Results from our own research indicate that EEG activations for monologue and dialogue listening have comparable spatial and temporal characteristics (Ip et al., in preparation)[1] One noteworthy phenomenon was an increased cortical tracking for dialogue listening, possibly due to increased listening engagement. That study also examined the impact of dysfluencies, an important element of speech communication that have been associated with multiple roles, such as bringing attentional focus to upcoming speech material (Diachek & Brown-Schmidt, 2023). However, dysfluencies may also pose a technical challenge. For example, lexical processing can be measured by relating neural signals with LLM-derived features (Chalehchaleh et al., 2025; Heilbron et al., 2022; Tikochinski et al., 2023). However, LLMs have been suggested to be impacted by dysfluencies, requiring tailored fine-tuning strategies (Mousavi et al., 2024). Our study indicates that, instead, the cortical tracking of speech can be robustly measured using LLM-derived features, using an off-the-shelf model, producing results compatible with monologue listening tasks (Ip et al., in preparation). Further advances are expected, given the growing interest and availability of publicly available neural data, such as the Podcast ECoG dataset, which offers intracranial recordings from nine individuals listening to podcasts, with the corresponding acoustic and linguistic features (Zada, Nastase, Aubrey, et al., 2025). This and other similar contributions producing precious data (Wang et al., 2025) can be very important, as that they share rare neural recordings that can be reanalysed in numerous manners, which could take years for a single research team.

Numerous other questions relevant to speech interaction can be explored with relatively controlled tasks involving the listening of **prerecorded monologue or dialogue**. For example, hypotheses such as the "**herding effect**", which proposes that a stronger speaker-listener alignment promotes how much the listeners cluster together (Chang et al., 2024), have been preliminarily tested using controlled listening paradigms involving pre-recorded speech. Let's now take communication accommodation as an example. Our recent work has investigated communication **accommodation** by comparing neural signatures of speech listening across different registers, such as native-directed and foreigner-directed speech (Piazza et al., 2025). We found that foreigner-directed speech has a pervasive impact on the neural processing of speech, aiding sound, phonological, and lexical encoding, beyond what could be achieved by simply speaking more slowly. One limitation of that work is that communication and accommodation are naturally two-directional phenomena (albeit with varying degrees of accommodation).

The next step brings us to experiments involving social interaction (i.e., interaction complexity in **Figure 3**). **Two-party (dyadic) conversations** have been used to study how neural activity aligns between two interlocutors. A variety of recording technologies have been explored for this research direction, including EEG (Solanki, 2017), MEG (Lin et al., 2023), fMRI (Misaki et al., 2021; Montague et al., 2002), fNIRS (Aili et al., 2025; Green et al., 2024; Li et al., 2024; Liu et al., 2019), and even intracranial EEG (Zada et al., 2024). The study of brain-to-brain coupling has been at the forefront of this line of research, proposing that interpersonal brain coupling arising via various means (e.g., body movement, breathing, focussing on the same stimulus) impacts the actions of an individual, generating behaviour that could not have emerged in isolation (Hasson & Frith, 2016; Hasson et al., 2012; Jiang et al., 2020; Pérez et al., 2017b). In line with that idea, previous research has studied social interaction with a focus on brain synchronisation across various social contexts, such as classroom learning (Dikker et al., 2017). Brain-to-brain synchrony has been related to similar communicative effort (Green et al., 2024), comprehension difficulties (Li et al., 2024), persuasion and deception (Angioletti et al., 2024; Veneziani et al., 2025), socio-psychological connectedness (Hinvest et al., 2025), delay in phone conversations (Uhrig et al., 2018), and other factors. One interesting application of these metrics is on the study of parent-child brain-to-brain synchrony, which has been investigated with multiple recording modalities across several ages and countries (Lin et al., 2023; Neel et al., 2025; Schwartz et al., 2024; Zhou et al., 2025). An issue with those metrics is that they may reflect the simultaneous alignment at a variety of levels, without pinpointing any of them in particular, unless a specific control condition is included.

---

[1] *Expected preprint publication date: June 2025. The reference will be added at the revision stage.*

Furthermore, recent work has used more targeted analytical methods to assess how language encoding is affected by social interaction. This phenomenon has been studied in dyads with fMRI (with participants being tested in neighbouring rooms), finding that strangers converge linguistically and neurally, while friends diverge over time (Speer et al., 2024). This result is compatible with the idea that strangers naturally search for a common ground, while friends explore more diverse topics. Other results on dyadic speech interaction using fMRI found that shared representations for listening and production extend beyond the language network into areas that are typically associated with social cognition (Zada, Nastase, Speer, et al., 2025). Further research was conducted to study what exactly drives the brain alignment during dyadic communication, finding that interlocutors align their neural representation of semantics (Zada et al., 2024) and syntax (Liu et al., 2019). Research using intracranial electrical recordings could go even deeper, finding that contextual embeddings learned by LLMs capture word-by-word neural alignment between speaker and listener (Zada et al., 2024). That research found, as expected, that the speaker's brain encodes linguistic content before word articulation, and that the same content re-emerges in the listener's brain after word articulation, providing precise spatial and temporal maps for that encoding. These findings were obtained despite dealing with the numerous technical complexities of dyadic interaction, supporting the feasibility and value of that experimental paradigm.

Indeed, there is so much more that we could talk about in the context of social interaction, such as multi-party interaction (Zhou et al., 2025), turn-taking (Bögels, 2020), non-verbal cues (Argyle, 1972) and even social phenomena beyond speech interaction (e.g., dyadic dance) (Bigand et al., 2024). Eye-gaze, for example, plays a major role in human communication and was suggested to be instrumental even in infants for bringing brains into mutual temporal alignment, facilitating information transfer and learning (Leong et al., 2017).

### 4.3. Future outlook

The previous section illustrates that the study of speech neurophysiology during realistic interaction is feasible. However, numerous challenges have yet to be addressed, both technical and theoretical, with some considerable differences across the social dimension (**Figure 1,** complexity of interaction). Nonetheless, the rising interest in this domain, exemplified through research studies (see previous section) and dedicated workshops as fora of discussion (e.g., TRF workshop 2024[2]; CNSP hackathon 2025[3]), gives us confidence that technical solutions will be identified as we move forward with this work. For argument's sake, let's project ourselves into a future where the technical barriers to social speech neurophysiology are solved, and running these experiments becomes well-defined and streamlined. Even with all technical barriers solved, we expect these experiments to remain more challenging than simple audiobook listening tasks. As such, identifying clear and important objectives that motivate this line of work, as well as developing unifying frameworks for investigation across this multi-disciplinary landscape, is crucial for ensuring its success, tailoring its developments toward the most important research questions. As such, it is crucial that ask ourselves the question: What is it that we are trying to solve? How is this line of work going to advance our understanding of the human brain?

Measurements of neural entrainment, stimulus-brain and brain-to-brain, have already been explored in this area. As such, we expect more of that work in the near future, with potential for substantially advancing our understanding of the link between neural entrainment mechanisms and social interaction. The possibility of studying brain-to-brain synchronisation in scenarios involving interaction adds an important dimension to this line of work. It will also be important to determine how the social element of speech impacts phenomena that are already studied in more controlled scenarios such as audiobook listening, from how speech and language features are encoded in the human brain to auditory attention mechanisms. While it is unclear if that work will constitute a breakthrough or simply lead to interesting complementary insights into brain functioning, we expect that studying the social dimension will offer new opportunities for enabling speech neurophysiology research to contribute in new ways to the study of the human brain. Other phenomena that subserve social speech communication (e.g., accommodation, anxiety, humour, bias, and empathy) will now have the opportunity of being studied with neurophysiology, by employing paradigms that involve realistic speech interaction, and fine-grained neural measurements at multiple timescales (e.g., phoneme, word, phrase) (Keitel et al., 2018) in addition to the more common brain-to-brain synchronisation.

A parallel contribution of this research involves human-machine interaction. As we have reached the uncanny valley in many technology domains and we attempt to get past that, behavioural reports might be insufficient for clearly guide this transformation, and biosignals may provide complementary insights. One example from that area involves multimodal speech neuroprosthetics, which have already been proposed (Littlejohn et al., 2025; Metzger et al., 2023), and one of the challenges is to make that form of embodied communication as natural as possible. Other applications include interlocutor modelling in the context of personal assistant interactions, conversational agents, and other platforms of human-computer dialogue (Clark, Pantidi, et al., 2019; Cowan et al., 2015; Doyle et al., 2019; Paola R. Peña et al., 2023). The distinction of such interlocutors or audience roles, as previously mentioned, is one example that highlights the importance of future research in this line of work to move towards harmonising terminology and forming unifying frameworks of investigation across domains. Such efforts ensure a consistency of insights leading to our ability as a field to present truly novel insights and discoveries about functions in the brain, aligned with existing theories and methodologies. This is evidenced in the psycholinguistics work that has already accumulated decades of research on speech perception and production, and now in the transition from human-human interaction to human-computer interaction in conversational user interfaces (Clark, Doyle, et al., 2019; Cowan et al., 2015; Paola R. Peña et al., 2023; P. R. Peña et al., 2023).

---

[2] https://hanse-ias.de/veranstaltungen/veranstaltung-detail-uebersicht/event/493
[3] https://cnspworkshop.net/

While we have mentioned many potential impacts of social speech neurophysiology, our view is that the greatest breakthrough might involve the study of mechanisms such as communication accommodation, anxiety, bias, and other socially-relevant phenomena that have largely eluded speech neurophysiology research so far. That work might be transformative for studying how those phenomena develop across the lifespan, and how they differ in particular cohorts (e.g., with deficits impacting social communication). So, big hype or big leap? It's probably a bit of both. The complexity of social speech neurophysiology experiments calls for a clear definition of the specific objectives, guiding technological developments in this area with targeted experimental paradigms and controls, reducing an otherwise enormous space of possible factors that play a role during speech interaction, and reconciling parallel developments and emerging frameworks in related fields. The additional experimental challenges compared with audiobook listening tasks will make standardisation and data sharing even more important here, enabling re-analyses from multiple angles and enabling contributions from teams with limited resources (e.g., multiple EEG systems). This will ultimately accelerate the development of this exciting domain of research, bridging speech neurophysiology with neighbouring research areas, such as pragmatics, perspective taking, human-machine interaction, anxiety, humour, and numerous others.